\documentclass[epj]{svjour}
\usepackage{amssymb}
\usepackage{graphicx}
\usepackage{dcolumn,color}
\usepackage{bm}
\usepackage{epstopdf}
\usepackage{epsfig}
\usepackage{color,framed}
\usepackage{amsmath,amssymb}
\begin{document}
\title{Localization of five-dimensional Elko spinors with non-minimal coupling on thick branes}

\author{Xiang-Nan Zhou$^{1}$,
        Yun-Zhi Du$^{2}$,
       Zhen-Hua Zhao$^{3}$,
      and Yu-Xiao Liu$^{4}$}                     
%
\mail{liuyx@lzu.edu.cn}
\institute{\small$^1$College of Physics and Information Engineering, Shanxi Normal University, Linfen 041004, People's Republic of China\\
  \small$^2$Institute of Theoretical Physics, Datong University, Datong 037009, People's Republic of China\\
  \small$^3$Department of Applied Physics, Shandong University of Science and Technology, Qingdao 266590, People's Republic of China\\
  \small$^4$Institute of Theoretical Physics, Lanzhou University, Lanzhou 730000, People's Republic of China}
\date{Received: date / Revised version: date}
%
\abstract{It has been found that the zero mode of a five-dimensional Elko spinor could be localized on branes by introducing a Yukawa-type
coupling between the Elko spinor and the background scalar field or the Ricci scalar. However, the Yukawa-type coupling is not appropriate for all brane models. In this paper, we explore other localization mechanism for the Elko spinor by introducing the non-minimal coupling $f(\phi)\mathfrak{L}_{Elko}$ between the five-dimensional Elko spinor and the background scalar field. We give the general expressions of the Elko zero mode and the function $f(\phi)$. Through two thick brane models and three concrete examples, we show that the Elko zero mode can be localized on the branes by this new mechanism. This provides us more possibilities of localizing the Elko zero mode.
} 
\maketitle
\section{Introduction}
\label{section1}
Brane-world models have attracted more and more interests since they were brought up
\cite{ArkaniHamed:1998rs,Randall:1999ee,Randall:1999vf}. They not only give a novel viewpoint of our world, but also open a new way to interpret many outstanding issues that the Standard Model (SM) can not interpret sufficiently, such as the hierarchy problem \cite{ArkaniHamed:1998rs,Randall:1999ee,Antoniadis:1998ig,Das:2007qn,Yang:2012dd,Guo:2015wua}, cosmological constant problem \cite{ArkaniHamed:2000eg,Kim:2000mc,Starkman:2000dy,Starkman:2001xu,Kehagias:2004fb,George:2008vu,Dey:2009xu,Haghani:2012zq}, the nature of dark matter and dark energy \cite{ArkaniHamed:1998nn,Sahni:2002dx,Cembranos:2003mr,Nihei:2004xv} and so on. Most of the early brane world models like the famous Randall-Sandrum (RS) branes were thin ones \cite{ArkaniHamed:1998rs,Randall:1999ee,Randall:1999vf} neglecting the thicknesses of branes. Subsequently, some more realistic branes with thickness (i.e., thick brane models) were proposed, whose energy densities have a distribution along the extra dimension. Usually, a thick brane can be generated dynamically by bulk
matter fields \cite{Gremm:1999pj,DeWolfe:1999cp,Kobayashi:2001jd,Bazeia:2002sd,Bazeia:2004dh,Afonso:2006gi,Bazeia:2006ef,Bogdanos:2006qw,Dzhunushaliev:2009va,Liu:2009ega,Liu:2011wi,Guo:2011wr,Liu:2012rc,Liu:2012gv,Bazeia:2012br,German:2013sk,Dutra:2014xla,Guo:2014bxa,Geng:2015kvs}, or by pure gravity \cite{BarbosaCendejas:2005kn,HerreraAguilar:2010kt,Zhong:2015pta}. More details about various kinds of thick branes could be found in Refs. \cite{Dzhunushaliev:2009va,Liu:2017}.

In brane world scenarios, one of the most important and interesting questions is to investigate the localization mechanism of various matter fields in a higher-dimensional spacetime on a 3+1 brane. The investigation of the Kaluza-Klein (KK) modes of various bulk matter fields can give us the way to probe the extra dimensions by considering their coupling with particles on the brane and the corrections to Newton's law and Coulomb's law \cite{Randall:1999vf,Hoyle:2000cv,Hung:2003cj,Adelberger:2006dh,Guo:2011qt}. In general, the zero modes of all five-dimensional matter fields, which correspond to the four-dimensional massless particles on a brane, must be localized on the brane to make sure the four-dimensional SM can be rebuilt on the brane at least at low energy. There were many works that studied the localization of various matter fields on branes  \cite{Guo:2011qt,Bajc:1999mh,Oda:2000zc,Liu:2007ku,Liu:2008wd,Liu:2009uca,Guerrero:2009ac,Zhao:2010mk,Chumbes:2010xg,Chumbes:2011zt,Germani:2011cv,Cruz:2012kd,Fu:2013ita,Liu:2013kxz,Zhao:2014gka,Alencar:2014moa,Vaquera-Araujo:2014tia,Fu:2015cfa,Fu:2016vaj,Zhang:2016ksq,Li:2017dkw}.

The localization of fields beyond SM have also been investigated. Elko spinor field, which was proposed by Ahluwalia and Grumiller \cite{Ahluwalia:2004sz,Ahluwalia:2004ab} in 2005, is a new quantum field with spin-1/2 beyond the four-dimensional SM and is the eigenspinor of the charge conjugation operator. However, Elko spinor field has some different properties compared with Dirac spinor: i) it satisfies the Klein-Gordon (KG) equation rather than the Dirac equation, ii) for the five-dimensional Elko spinor, its mass dimension is one, instead of 3/2 that is suit for Dirac spinor. Hence the interactions between Elko spinor and gauge fields in SM will be suppressed strongly (at least one order of Plank scale), which means that Elko spinor only interacts with itself, graviton and Higgs fields \cite{Ahluwalia:2004sz,Ahluwalia:2004ab,Ahluwalia:2008xi,Ahluwalia:2009rh,Ahluwalia:2010zn,Dias:2010aa,Lee:2012td,Lee:2015jpa,Fabbri:2010va}. Elko spinor has been considered as a first-principle candidate of dark matter \cite{Ahluwalia:2004sz,Ahluwalia:2004ab,Ahluwalia:2008xi,Ahluwalia:2009rh,Ahluwalia:2010zn} and has also been investigated extensively in cosmology \cite{Boehmer:2008rz,Boehmer:2008ah,Boehmer:2010ma,Wei:2010ad,Gredat:2008qf,Basak:2014qea,Pereira:2014wta,Pereira:2016emd,Pereira:2017efk,Pereira:2017bvq} and mathematical physics \cite{HoffdaSilva:2009is,HoffdaSilva:2016ffx,daRocha:2011yr,Ablamowicz:2014rpa,Rogerio:2016grn,Cavalcanti:2014wia,Cavalcanti:2014uta}.

The localization of Elko spinor in five-dimensional brane world models has been investigated in Refs. \cite{Liu:2011nb,Jardim:2014cya,Jardim:2014xla}. It was found that the zero mode of Elko spinor can be localized on branes thought a Yukawa-type coupling with the background scalar field \cite{Liu:2011nb,Jardim:2014xla}, or
the Ricci scalar \cite{Jardim:2014xla}. However, the Yukawa-type coupling is only appropriate for some kinds of thick branes, and the coupling constant must be taken as some particular expression determined by the parameters in the brane models. This motivates us to consider the other kinds of couplings to release the limits of coupling constants and brane models. In Ref.~\cite{Liu:2017}, a general coupling between the five-dimensional Dirac fermion and the background scalar fields is introduced, which includes the non-minimal coupling between the Dirac fermion and the scalar fields. Inspired by this, in this paper, we will introduce this new type of coupling, i.e., the non-minimal coupling $f(\phi)\mathfrak{L}_{Elko}$ between the five-dimensional Elko spinor and the background scalar field, where $f(\phi)$ is a function of the backgound scalar field and $\mathfrak{L}_{Elko}$ is the Lagrangian of the five-dimensional Elko spinor. We will study the localization of the Elko spinor with this new coupling on the Minkowski thick branes which have been considered in Yukawa-type coupling case \cite{Liu:2011nb} and compare the characteristics between the Yukawa-type coupling and this new coupling. We will show that localized Elko zero mode on the branes can be realized for different forms of $f(\phi)$, which means that this new coupling can provide us more possibilities of localizing the Elko zero mode. By the way, the localization of the Elko spinor in a six-dimensional string-like model was considered in Ref. \cite{Dantas:2015mfi}, recently.

The organization of the paper is listed as follows. In Sec.~\ref{section2}, we give a briefly review of Elko spinor field. In Sec.~\ref{section2}, we consider the localization of a five-dimensional massless Elko spinor with a non-minimal coupling to the background scalar field and give the equation of the Elko zero mode. Then in Sec.~\ref{section3}, we discuss the localization of the Elko zero mode on various thick branes. Finally, the conclusion is given in Sec.~\ref{section4}.

\section{Review of Elko spinor field}
\label{section2}
Elko spinor field is a spin-1/2 matter field, but it is very different from the Dirac fermion spinor. Elko spinor field belongs to non-standard Wigner classes and can not be expressed in Weinberg's formalism \cite{Ahluwalia:2004sz,Ahluwalia:2010zn}. Elko spniors satisfy the unusual property $(CPT)^2=-\mathbb{I}$, where the charge conjugation $C$ {is} defined as
\begin{eqnarray}
C=\left( \begin{array}{cc}
\mathbb{O}   & i\Theta \\
-i\Theta     & \mathbb{O}
\end{array} \right)K.
\end{eqnarray}
Here $K$ is the complex conjugation operator, and $\Theta$ is the spin one half Wigner time reversal operator which satisfies $\Theta(\vec{\sigma}/2)\Theta^{-1}=-(\vec{\sigma}/2)^{*}$.
Elko spinors are eigenspinors of the charge conjugation operator: $C\lambda(k^\mu)=\pm\lambda(k^\mu)$ with $k^\mu$ a polarization vectorx. The plus sign generates the self-conjugate spinors which are denoted by $\varsigma(k^\mu)$ and the minus sign generates the anti-self-conjugate spinors which are denoted by $\tau(k^\mu)$. Both of the two kinds of spinors have two possible helicities $\chi_{\pm}(k^\mu)$, which can be expressed as
\begin{eqnarray}
\chi_{+}(k^{\mu})=e^{-\text{i}\phi/2}\sqrt{m}{1 \choose 0},~~\chi_{-}(k^{\mu})=e^{\text{i}\phi/2}\sqrt{m}{0 \choose 1}.~~\label{chim}
\end{eqnarray}
Thus there exist four types of Elko spinors which are written as
\begin{eqnarray}
\varsigma_\pm(k^\mu) &=& ~~{\text{i}\Theta[\chi_{\pm}(k^{\mu})]^{*} \choose \chi_\pm(k^{\mu})}, \\
\tau_\pm(k^\mu) &=&\pm{-\text{i}\Theta[\chi_{\mp}(k^{\mu})]^{*} \choose \chi_\mp(k^{\mu})}.~~~~
\end{eqnarray}

{
By using a transformation operator $\Gamma$, $k^{\mu}$ can be transformed as a general vector $p^{\mu}$ which represents ($E$, $p_x$, $p_y$, $p_z$). Here the $\Gamma$ is given by \cite{Ahluwalia:2010zn}
\begin{eqnarray}
\Gamma=\left( \begin{array}{cccc}
\sqrt{\frac{m}{E-p_z}} & \frac{p_x-\text{i}p_y}{\sqrt{m(E-p_z)}} & 0                                         & 0 \\
                 0     & \sqrt{\frac{E-p_z}{m}}                  & 0                                         & 0 \\
                 0     &                  0                      & \sqrt{\frac{E-p_z}{m}}                    & 0 \\
                 0     &                  0                      & -\frac{p_x+\text{i}p_y}{\sqrt{m(E-p_z)}}  & \sqrt{\frac{m}{E-p_z}}
\end{array} \right).~~~
\end{eqnarray}
Then by taking the massless limit in the expression of Elko spinor $\lambda(p^{\mu})$, the four-dimensional massless Elko spinor can be got. We can find that the $\varsigma_{-}(p^\mu)$ and $\tau_{+}(p^\mu)$ will vanish but $\varsigma_{+}(p^\mu)$ and $\tau_{-}(p^\mu)$ will not in the massless limit.} In addition, the dual spinors of Elko spinors are defined as
\begin{eqnarray}
  {\mathop {\varsigma}\limits^\neg}_{\pm}(p^{\mu})=\pm[\varsigma_{\mp}(p^{\mu})]^{\dag}\gamma^{0},~~
  {\mathop {\tau}\limits^\neg}_{\pm}(p^{\mu})=\pm[\tau_{\mp}(p^{\mu})]^{\dag}\gamma^{0}.
\end{eqnarray}

Elko spinors can not be expressed in Weinberg's formalism. Thus they do not satisfy the usual Dirac equation. By using the Dirac operator $\gamma_{\mu}p^{\mu}$ to act on Elko spinors, one can get
\begin{eqnarray}
  \gamma_{\mu}p^{\mu}\varsigma_{\pm}(p^{\mu})=\mp m\varsigma_{\mp}(p^{\mu}),~
  \gamma_{\mu}p^{\mu}\tau_{\pm}(p^{\mu})=\pm m\tau_{\mp}(p^{\mu}).~~~~
\end{eqnarray}
Here the gamma matrices $\gamma^\mu$ satisfy the relation $\{\gamma^\mu,\gamma^\nu\}=2\eta^{\mu\nu}\mathbb{I}$ with $\eta^{\mu\nu}=\text{diag}(-,+,+,+)$. At the same time, when $\gamma^5$ acts on Elko spinors, the results are
\begin{eqnarray}
\gamma^5\varsigma_{\pm}(p^{\mu})=\pm\tau_{\mp}(p^\mu),\qquad \gamma^5\tau_{\pm}(p^\mu)=\mp\varsigma_\mp(p^{\mu}).
\end{eqnarray}
Then the following equations can be obtained by the Fourier transformation:
\begin{eqnarray}
\gamma^\mu\partial_\mu\varsigma_\pm(x)&=&\mp \text{i}m\varsigma_\mp(x),~\gamma^\mu\partial_\mu\tau_\pm(x)=\pm \text{i}m\tau_\mp(x); ~~~~ \label{5D Dirac operator on Elko spinors}\\
\gamma^5\varsigma_{\pm}(x)&=&\pm\tau_{\mp}(x),\qquad~~\gamma^5\tau_{\pm}(x)=\mp\varsigma_\mp(x). \label{5D gamma 5 on Elko spinors}
\end{eqnarray}
It should be noticed that it is $\gamma^{\mu}\partial_\mu\psi\propto\psi$, where $\psi$ is a usual Dirac spinor. Although Elko spinors do not satisfy the usual Dirac equation, they satisfy the KG equation: $(\eta^{\mu\nu}\partial_\mu\partial_\nu-m^{2})\lambda(x)=0$. Thus, the Lagrangian density of a free Elko in four-dimensional flat space-time reads $\mathfrak{L}_{\text{Elko}}=-\frac{1}{2}\partial^{\mu}\mathop\lambda\limits^\neg\partial_{\mu}\lambda-\frac{1}{2}m^{2}\mathop \lambda\limits^\neg\lambda$. And for a general curved space-time, the Lagrangian for a free massless Elko should read \cite{Ahluwalia:2004ab,Gredat:2008qf,Wei:2010ad}
\begin{eqnarray}
  \mathfrak{L}_{\text{Elko}}=-\frac{1}{4}g^{\mu\nu}(\mathfrak{D}_{\mu}
  \mathop \lambda\limits^\neg\mathfrak{D}_{\nu}\lambda+\mathfrak{D}_{\nu}
  \mathop \lambda\limits^\neg\mathfrak{D}_{\mu}\lambda),
  \label{Elko's Lagrangian density}
\end{eqnarray}
where
$\mathfrak{D}_{\mu}$ represents covariant derivative.

\section{Localization of five-dimensional Elko spinor with non-minimal coupling}
\label{section3}
In order to localize a five-dimensional Elko spinor on a brane with codimension one, a Yukawa-type coupling between the Elko spinor and the background scalar field or the scalar curvature $R$ has been considered in Refs.~\cite{Liu:2011nb,Jardim:2014xla}. However, the Yukawa-type coupling is not appropriate for all brane models and the expression of the coupling constant will be determined exactly by the parameters in the models. Therefore, we will focus on the non-minimal coupling between a five-dimensional Elko spinor and the background scalar field in this section. The five-dimensional line-element is assumed as
\begin{eqnarray}
ds^{2}=\text{e}^{2A(y)}\hat{g}_{\mu\nu}dx^{\mu}dx^{\nu}+dy^{2}. \label{the line-element for 5D space time}
\end{eqnarray}
Here, $\text{e}^{2A(y)}$ is the warp factor and $\hat{g}_{\mu\nu}$ is the induced metric on the brane. Performing the following coordinate transformation
\begin{eqnarray}
dz=\text{e}^{-A(y)}dy, \label{coordinate transformation}
\end{eqnarray}
the metric (\ref{the line-element for 5D space time}) can transform to be
\begin{eqnarray}
ds^{2}=\text{e}^{2A}(\hat{g}_{\mu\nu}dx^{\mu}dx^{\nu}+dz^{2}), \label{conformally flat line-element}
\end{eqnarray}
which is more convenient for discussing the localization of various matter fields on branes. As the same as the localization mechanism of the Dirac fermion field on brane considered in \cite{Liu:2017}, the action of a five-dimensional massless Elko spinor with non-minimal coupling to background scalars is written as
\begin{eqnarray}
S&=&\int d^5x\sqrt{-g}f(\phi)\mathfrak{L}_{Elko},\nonumber \\
\mathfrak{L}_{\text{Elko}}&=&-\frac{1}{4}g^{MN}\left(\mathfrak{D}_{M}\mathop \lambda\limits^\neg\mathfrak{D}_{N}\lambda+\mathfrak{D}_{N}\mathop \lambda\limits^\neg\mathfrak{D}_{M}\lambda\right).~~~~\label{Lagrangian for Elko in 5D}
\end{eqnarray}
Here $f(\phi)$ is a function of the background scalar field $\phi$ and $\phi$ is a function of the extra dimension. $\lambda$ represents a five-dimensional Elko field and $\mathop \lambda\limits^\neg$ is the dual one. The capital Latin letters ($M,~N=0,1,2,3,5$) and the Greek ones ($\mu,~\nu=0,1,2,3$) denote the five-dimensional and four-dimensional space-time indices, respectively. In addition, we let $\bar{A},\bar{B}\cdots=0,1,2,3,5$ and $a,b\cdots=0,1,2,3$ denote the five-dimensional and four-dimensional local Lorentz indices, respectively. The covariant derivatives are
\begin{eqnarray}
\mathfrak{D}_{M}\lambda=(\partial_{M}+\Omega_{M})\lambda,~~\mathfrak{D}_{M}\mathop \lambda\limits^\neg=\partial_{M}\mathop \lambda\limits^\neg-\mathop \lambda\limits^\neg\Omega_{M}. \label{covariant derivatives}
\end{eqnarray}
Here the spin connection $\Omega_{M}$ is given by
\begin{eqnarray}
   \Omega_{M}&=&-\frac{i}{2}\left(e_{\bar{A}P}e_{\bar{B}}^{~~N} \Gamma^{P}_{MN}
                             -e_{\bar{B}}^{~~N}\partial_{M}e_{\bar{A}N}\right)
                  S^{\bar{A}\bar{B}},\nonumber \\
   S^{\bar{A}\bar{B}}&=&\frac{i}{4}[\gamma^{\bar{A}},\gamma^{\bar{B}}],
    \label{tangent space connection}
\end{eqnarray}
where $e^{\bar{A}}_{~M}$ is the vierbein and satisfies the orthonormality relation $g_{MN}=e^{\bar{A}}_{~M}e^{\bar{B}}_{~N}\eta_{\bar{A}\bar{B}}$. So the non-vanishing components of the spin connection $\Omega_{M}$ for the metric \eqref{conformally flat line-element} are:
\begin{eqnarray}
\Omega_{\mu}=\frac{1}{2}\partial_{z}A\gamma_{\mu}\gamma_{5} +  \hat{\Omega}_{\mu}, \label{spin connection for brane}
\end{eqnarray}
where the $\gamma_{\mu}$ and $\gamma_{5}$ are the four-dimensional gamma matrixes on the brane and the $\gamma_{\mu}$ satisfy $\{\gamma_{\mu},\gamma_{\nu}\}=2\hat{g}_{\mu\nu}$, and $\hat{\Omega}_{\mu}$ is the the spin connection on the brane.

The equation of motion for the Elko can be got from the action (\ref{Lagrangian for Elko in 5D})
\begin{eqnarray}
\frac{1}{\sqrt{-g}f(\phi)}\mathfrak{D}_{M}(\sqrt{-g}f(\phi)g^{MN}\mathfrak{D}_{N}\lambda)=0. \label{Elko's motion equation}
\end{eqnarray}
With the conformally metric (\ref{conformally flat line-element}) and the non-vanishing components of the spin connection (\ref{spin connection for brane}), the above equation (\ref{Elko's motion equation}) becomes
\begin{eqnarray}
  \frac{1}{\sqrt{-\hat{g}}}\hat{\mathfrak{D}}_{\mu}(\sqrt{-\hat{g}}\hat{g}^{\mu\nu}\hat{\mathfrak{D}}_{\nu}\lambda)
 +\bigg[-\frac{1}{4}{A'}^{2}\hat{g}^{\mu\nu}\gamma_{\mu}\gamma_{\nu}\lambda\nonumber\\
 +\frac{1}{2}A'
 \Big( \hat{\mathfrak{D}}_{\mu}
       (  \hat{g}^{\mu\nu}\gamma_{\nu}\gamma_{5}\lambda)
 +\hat{g}^{\mu\nu}\gamma_{\mu}\gamma_{5}\hat{\mathfrak{D}}_{\nu}\lambda\Big)
 \nonumber\\
 +\text{e}^{-3A}f^{-1}(\phi)\partial_{z}(\text{e}^{3A}f(\phi)\partial_{z}\lambda)\bigg]=0,
  \label{ElkoMotionEq2}
\end{eqnarray}
where $\hat{\mathfrak{D}}_{\mu}$ denotes the four-dimensional covariant derivatives on the brane and it satisfies $\hat{\mathfrak{D}}_{\mu}\lambda=(\partial_{\mu}+\hat{\Omega}_{\mu})\lambda$.
With $\hat{\mathfrak{D}}_{\mu}\hat{e}^{a}_{~\nu}=0$, one has  $\hat{\mathfrak{D}}_{\mu}\hat{g}^{\lambda\rho}=\hat{\mathfrak{D}}_{\mu}(\hat{e}_{a}^{~\lambda}\hat{e}^{a\rho})=0$. Thus, the above equation can be simplified as
\begin{eqnarray}
 \frac{1}{\sqrt{-\hat{g}}}\hat{\mathfrak{D}}_{\mu}(\sqrt{-\hat{g}}\hat{g}^{\mu\nu}\hat{\mathfrak{D}}_{\nu}\lambda)
  -\! A'\gamma^{5}\gamma^{\mu} \hat{\mathfrak{D}}_{\mu}\lambda
   \!-\! {A'}^{2}\lambda \nonumber\\
   \!+\!\text{e}^{-3A}f^{-1}(\phi)\partial_{z}(\text{e}^{3A}f(\phi)\partial_{z}\lambda)=0.~~
  \label{Elko's motion equation for flat brane}
\end{eqnarray}
As what we have done in our previous work~\cite{Liu:2011nb,Jardim:2014xla}, we introduce the KK decomposition of the five-dimensional Elko spinor
\begin{eqnarray}
\lambda_{\pm}&=&\text{e}^{-3A/2}f(\phi)^{-\frac{1}{2}}\sum_{n}\left(\alpha_{n}(z)\varsigma^{(n)}_{\pm}(x)+\alpha_{n}(z)\tau^{(n)}_{\pm}(x)\right)\nonumber\\
             &=&\text{e}^{-3A/2}f(\phi)^{-\frac{1}{2}}\sum_{n}\alpha_{n}(z)\hat{\lambda}_{\pm}^{n}(x).\label{Elkodecomposition}
\end{eqnarray}
Here and after we omit the $\pm$ subscript of the $\alpha_n$ functions for simplicity. The functions $\varsigma^{n}_{\pm}(x)$ and $\tau^{(n)}_{\pm}(x)$ are four-dimensional linear independent Elko spinors, which satisfy
\begin{eqnarray}
\gamma^{\mu}\hat{\mathfrak{D}}_{\mu}\varsigma_{\pm}(x)=\mp\text{i}\varsigma_{\mp}(x),~~~\gamma^{\mu}\hat{\mathfrak{D}}_{\mu}\tau_{\pm}(x)&=&\pm\text{i}\tau_{\mp}(x),\\
\gamma^5\varsigma_{\pm}(x)=\pm\tau_{\mp}(x),~~~~~~~~\gamma^5\tau_{\pm}(x)&=&\mp\varsigma_{\mp}(x).
\end{eqnarray}
At the same time the four-dimensional Elko spinor should satisfy the K-G equation
\begin{eqnarray}
\frac{1}{\sqrt{-\hat{g}}}\hat{\mathfrak{D}}_{\mu}(\sqrt{-\hat{g}}\hat{g}^{\mu\nu}\hat{\mathfrak{D}}_{\nu}\lambda_n)=m^2_n\lambda_n
\end{eqnarray}
with $m_n$ the mass of the Elko spinor on the brane. Therefore, the equation of the KK mode $\alpha_{n}$ can be got
\begin{eqnarray}
\alpha_{n}''-\bigg[&-&\frac{1}{4}f^{-2}(\phi)f'^{2}(\phi)+\frac{3}{2}A'f^{-1}(\phi)f'(\phi)+\frac{1}{2}f^{-1}(\phi)f''(\phi)\nonumber\\
&+&\frac{3}{2}A''+\frac{13}{4}(A')^{2}-m_{n}^2+\text{i}m_{n}A'\bigg]\alpha_{n}=0.\label{KKequationforElko}
\end{eqnarray}
By introducing the following orthonormality conditions for $\alpha_{n}$
\begin{eqnarray}
\int \alpha^{*}_{n}\alpha_{m}dz&=&\delta_{nm},\label{orthonormality relation 1}
\end{eqnarray}
we can get the action of the four-dimensional massless and massive Elko spinors from the action (\ref{Lagrangian for Elko in 5D})
\begin{eqnarray}
S&=&-\frac{1}{4}\int d^5x\sqrt{-g}f(\phi)g^{MN}(\mathfrak{D}_M\mathop \lambda\limits^\neg\mathfrak{D}_{N}\lambda+\mathfrak{D}_N\mathop \lambda\limits^\neg\mathfrak{D}_{M}\lambda)\nonumber\\
               &=&-\frac{1}{2}\sum_{n}\int d^4x(\partial^{\mu}\hat{{\mathop \lambda\limits^\neg}}_{n}\partial_{\mu}\hat{\lambda}_{n}+m_{n}^2\hat{{\mathop \lambda\limits^\neg}}_{n}\hat{\lambda}_{n}),
\end{eqnarray}
where $\hat{\lambda}_{n}$ are the four-dimensional Elko spinors.

It is obvious that this term $\text{i}m_nA'$ in Eq. (\ref{KKequationforElko}) will bring us difficulty to investigate the massive Elko KK modes, which will be discussed in our future work. In this paper we just focus on the localization of the zero mode, which denotes the four-dimensional massless Elko spinor on the brane (i.e., $m_n=0$). Thus, Eq. (\ref{KKequationforElko}) for the massless case $m_0=0$ can be simplified as
\begin{eqnarray}
[-\partial_{z}^{2}+V_{0}(z)]\alpha_{0}(z)=0, \label{KKequationforzeromode}
\end{eqnarray}
with the effective potential given by
\begin{eqnarray}
V_{0}(z)=&-&\frac{1}{4}\frac{f'^{2}(\phi)}{f^{2}(\phi)}+\frac{3}{2}A'\frac{f'(\phi)}{f(\phi)}
                      +\frac{1}{2}\frac{f''(\phi)}{f(\phi)}\nonumber\\
         &+&\frac{3}{2}A''+\frac{13}{4}A'^2. \label{effective potential Vz}
\end{eqnarray}
The orthonormality condition for the Elko zero mode is
\begin{eqnarray}
\int \alpha^*_0\alpha_0dz=1.\label{orthonormality relation for zero mode}
\end{eqnarray}
Since there exist three types of functions (the negative power and derivative of $f(\phi)$, the derivative of A) in the effective potential (\ref{effective potential Vz}), we define a new function $B(z)=\frac{f'(\phi(z))}{f(\phi(z))}$ to transform Eq.~(\ref{effective potential Vz}) as
\begin{eqnarray}
V_0(z)=\frac{1}{4}B^2+\frac{3}{2}A'B+\frac{1}{2}B'+\frac{3}{2}A''+\frac{13}{4}A'^2. \label{effective potential VzB}
\end{eqnarray}

From Ref.~\cite{Liu:2011nb}, it is know that there exist many similarities between the Elko field and scalar field. For a five-dimensional free massless scalar field, the Schr\"{o}dinger-like equation for the scalar field zero mode $h_0(z)$ reads as~\cite{Liu:2008wd}
\begin{eqnarray}
[-\partial_{z}^{2}+V_{\Phi}]h_0
&=&\left[-\partial_{z}^{2}+\frac{3}{2}A''+\frac{9}{4}{A'}^{2}\right]h_0\nonumber\\
&=&\left[\partial_{z}+\frac{3}{2}A'\right]\left[-\partial_{z}+\frac{3}{2}A'\right]h_{0}\nonumber\\
&=&0.
\end{eqnarray}
The scalar zero mode can be solved as
\begin{eqnarray}
h_{0}(z)\varpropto\text{e}^{\frac{3}{2}A(z)},
\end{eqnarray}
and it always satisfies the orthonormality relation for any brane embedded in a five-dimensional Anti-de Sitter (AdS) space-time. It is the result of the fact that the coefficient of $A'^2$ is the square of the coefficient of $A''$.
Come back to the case of the Elko spinor with non-minimal coupling, it is not difficult to find that Eq. (\ref{KKequationforzeromode}) can  be rewritten as
\begin{eqnarray}
&&[-\partial_{z}^{2}+V_{0}]\alpha_{0}\nonumber\\
&=& \left[-\partial_{z}^{2}+\frac{1}{4}B^2+\frac{3}{2}A'B+\frac{1}{2}B'+\frac{3}{2}A''+\frac{13}{4}A'^2\right]\alpha_{0}\nonumber\\
&=&\left[\partial_{z}+\left(\frac{3}{2}A'+\frac{1}{2}B\right)\right]\left[-\partial_{z}+\left(\frac{3}{2}A'+\frac{1}{2}B\right)\right]\alpha_{0}+A'^2\alpha_0\nonumber\\
&=&0.
\end{eqnarray}
Note that although the term $A'^2\alpha_0$ will make the trouble, the form of $B(z)$ is arbitrary, which means that we can choice an appropriate $B(z)$ to eliminate the term $A'^2\alpha_0$. To this end, we introduce two new functions $C(z)$ and $D(z)$ that satisfy the following equations
\begin{eqnarray}
D''+D'^2&=&\frac{1}{4}B^2+\frac{3}{2}A'B+\frac{1}{2}B'+\frac{3}{2}A''+\frac{13}{4}A'^2,\label{D1} \\
D'&=&\frac{3}{2} A'+\frac{1}{2} B+C.                                                            \label{D2}
\end{eqnarray}
Substituting Eq. (\ref{D2}) into Eq. (\ref{D1}), we can get the relation between $C(z)$ and $B(z)$
\begin{eqnarray}
B(z)=-3A'+\frac{A'^2}{C}-C-\frac{C'}{C}.
\end{eqnarray}
Therefore, Eq. (\ref{KKequationforzeromode}) can be written as
\begin{eqnarray}
[-\partial_{z}^{2}+V_{0}]\alpha_{0}
&=&[-\partial_{z}^{2}+D''+D'^2]\alpha_{0}\nonumber\\
&=&\left[\partial_{z}+D'\right]\left[-\partial_{z}+D'\right]\alpha_{0}\nonumber\\
&=&0,
\end{eqnarray}
and the zero mode of the Elko spinor can be solved as
\begin{eqnarray}
 \alpha_{0}(z)&\varpropto& \text{e}^{D(z)}\nonumber\\
 &=&{\exp} \left[\frac{1}{2}\int^{z}_{0}\left(\frac{A'^2}{C}-\frac{C'}{C}+ C \right)d\bar{z}\right]. \label{zeromodeforz}
\end{eqnarray}
In order to obtain the localized zero mode, the function $C(z)$ is required to be an odd one. We introduce
\begin{eqnarray}
K(z)=\frac{C'}{C}-C, \label{K}
\end{eqnarray}
which is an odd function and is positive as $z>0$. Thus the zero mode can be written as
\begin{eqnarray}
 \alpha_{0}(z) 
 &=&{\exp}\left[\frac{1}{2}\int^{z}_{0}\left(\frac{A'^2}{C}-K\right)d\bar{z}\right]\nonumber\\
 &=&{\exp}\left[\frac{1}{2}\int^{z}_{0}\frac{A'^2}{C}d\bar{z}\right]
    {\exp}\left[-\frac{1}{2}\int^{z}_{0}Kd\bar{z}\right].\label{zeromodeforz2}
\end{eqnarray}
And the function $f(\phi(z))$ is expressed as
\begin{eqnarray}
f(\phi(z))&=&\text{e}^{\int^{z}_{0} B(\bar{z}) d\bar{z}}\nonumber\\
          &=&{\exp}\left[\int^{z}_{0}\left( -3A'+\frac{A'^2}{C}- C-\frac{C'}{C} \right) d\bar{z}\right]\nonumber\\
          &=&{\exp}\left[\int^{z}_{0}\left( -3A'+\frac{A'^2}{C}-K-2\frac{C'}{C} \right) d\bar{z}\right]\nonumber\\
          &=&\text{e}^{-3A}C^{-2}{\exp}\left[\int^{z}_{0}\frac{A'^2}{C}d\bar{z}\right]{\exp}\left[\int^{z}_{0}Kd\bar{z}\right].~~~~\label{fphi}
\end{eqnarray}
It should be noticed that here the function $C(z)$ can be expressed by the function $K(z)$ according to Eq.~(\ref{K}): $C(z)=\frac{\text{e}^{\int^z_1 K(\bar{z})d\bar{z}}}{\text{constant}-\int_1^z\text{e}^{\int^{\hat{z}}_1 K(\bar{z})d\bar{z}}d\hat{z}}$. Thus Eqs.~(\ref{zeromodeforz2}) and (\ref{fphi}) give the general expressions of the zero mode $\alpha_0$ and the function $f(\phi)$. For a given $K(z)$ the zero mode $\alpha_0$ is obtained by integrating the equation (\ref{zeromodeforz2}). Then, the scalar field function $f(\phi(z))$ is determined by integrating the equation (\ref{fphi}). We will find that the role of $K(z)$ is similar to the auxiliary superpotential $W(\phi)$, which is introduced in order to solve the Einstein equations in thick brane models.
As the superpotential $W(\phi)$ does in thick brane models, different $K(z)$ could lead to different solutions. To illustrate this, next we will review two kinds of thick brane solutions and obtain the zero mode by considering different forms of $K(z)$, for example.

\section{Localization of the Elko zero mode with non-minimal coupling on thick branes }
\label{section4}
In this section, we consider two brane models as examples: one is constructed by a standard canonical scalar field \cite{Dzhunushaliev:2009va,Gremm:1999pj,Afonso:2006gi} and the other is generated by a scalar field non-minimally coupled to the Ricci scalar curvature \cite{Dzhunushaliev:2009va,Bogdanos:2006qw,Guo:2011wr,Liu:2012gv}. Both of them are Minkowski thick branes embedded in asymptotically AdS space-time and have been investigated in our previous work \cite{Liu:2011nb}. By considering the localization of the five-dimensional Elko spinor on these thick branes, we can compare the Yukawa-type coupling and non-minimal coupling. The thick brane action of a standard canonical scalar coupled to gravity can be written as
\begin{eqnarray}
S={\int}d^{5}x\sqrt{-g}\left[\frac{1}{2}R-\frac{1}{2}(\partial\phi)^{2}-V(\phi)\right].
\end{eqnarray}
By considering the above action and the Minkowski brane metric (which means $\hat{g}_{\mu\nu}=\eta_{\mu\nu}$ in \eqref{the line-element for 5D space time})
\begin{eqnarray}
ds^{2}=\text{e}^{2A(y)}\eta_{\mu\nu}dx^{\mu}dx^{\nu}+dy^{2}, \label{the line-element Minkowski_brane}
\end{eqnarray}
the Einstein and scalar field equations are given by
\begin{subequations}\label{Einstein and scalar equation}
\begin{eqnarray}
&&\frac{\partial V(\phi)}{\partial\phi}=\phi''+4A'\phi',\\
&&6A'^2=\frac{1}{2}\phi'^2-V(\phi),\\
&&A''=-\frac{1}{3}\phi'^2.
\end{eqnarray}
\end{subequations}
By introducing the auxiliary superpotential $W(\phi)$, the scalar field potential is given as
\begin{eqnarray}
V(\phi)=-6W(\phi)^2+\frac{9}{2}\left(\frac{\partial W(\phi)}{\partial\phi}\right)^2. \label{potentialV1}
\end{eqnarray}
And we have
\begin{eqnarray}
\phi'=3\frac{\partial W(\phi)}{\partial\phi},~~~~~~~~~~~A'=-W(\phi).
\end{eqnarray}
Thus, for a given superpotential $W(\phi)$, the scalar field and the warp factor can be obtained by integrating the above equations. We will find that the role of $K(z)$ is similar to the superpotential $W(\phi)$. By choosing a simple superpotential \cite{Dzhunushaliev:2009va}
\begin{eqnarray}
W(\phi)=c\sin(b\phi),
\end{eqnarray}
the corresponding scalar potential is a sine-Gordon one
\begin{eqnarray}
V(\phi)=\frac{3}{2}c^{2}[3b^{2}\cos^{2}(b\phi)-4\sin^{2}(b\phi)]
\end{eqnarray}
and the brane solution is given by~\cite{Dzhunushaliev:2009va,Gremm:1999pj,Afonso:2006gi}
\begin{subequations}\label{solution1}
\begin{eqnarray}
 \text{e}^{A(y)}&=&\left[{\cosh(cb^{2}y)}\right]^{-\frac{1}{3b^2}},  \\
 \phi(y)&=&\frac{2}{b}\arctan \tanh\big(\frac{3}{2}cb^{2}y\big),
\end{eqnarray}
\end{subequations}
where $b$ and $c$ are parameters related to the brane thickness. The potential approaches the negative values $V(\pm\infty)=-6c^2$. Thus the bulk is asymptotically AdS. For simplicity, we define the parameters $\frac{1}{3b^2}=\bar{b}$, $cb^2=a$. Then, the brane solution becomes
\begin{subequations}\label{solution2}
\begin{eqnarray}
A(y) &=&-\bar{b}\ln\cosh(a y),\\
\phi(y) &=& \phi_0\arctan\tanh(\frac{3ay}{2}),
\end{eqnarray}
\end{subequations}
where $\phi_0=2\sqrt{3\bar{b}}$. If we take the solution (\ref{solution1}) into to Eq.~(\ref{Einstein and scalar equation}), we will find that it is indeed the solution to the Einstein and scalar field equations.

In addition, the thick brane solution of a scalar field non-minimally coupled to the Ricci scalar curvature was discussed in Refs. \cite{Dzhunushaliev:2009va,Bogdanos:2006qw,Liu:2012gv}. The action can be written as
\begin{eqnarray}
S=\int d^{5}x\sqrt{-g}\left[F(\phi)R-\frac{1}{2}(\partial\phi)^{2}-V(\phi)\right],
\end{eqnarray}
where $F(\phi)$ is a function of the scalar field $\phi$. By the conformal transformation $g_{MN}\rightarrow 2\widetilde{g}_{MN} F(\phi)$, the action is conformally corresponded to the Einstein frame one. For the Minkowski brane metric and the coupling function $
F(\phi)=\frac{1}{2}(1-\xi\phi^{2})$ with $\xi\neq0$, the Einstein equations are given by
\begin{subequations}\label{Einstein and scalar equation2}
\begin{eqnarray}
V(\phi)&=&-\frac{3}{2}\left(\frac{1}{2}-\frac{1}{2}\xi\phi^2\right)(2A'^2+A'')\nonumber\\
        &&+\frac{7}{2}\xi A'\phi'\phi+\xi\phi''\phi+\xi\phi'^2,\\
\frac{1}{2}\phi'^2&=&-\frac{3}{2}\left(\frac{1}{2}-\frac{1}{2}\xi\phi^2\right)A''\nonumber\\
&&-\frac{1}{2}\xi A'\phi'\phi+\xi\phi''\phi+\xi\phi'^2.
\end{eqnarray}
\end{subequations}
By choosing the warp factor as
\begin{eqnarray}
\text{e}^{A(y)}&=&\big[\cosh(ay)\big]^{-\gamma},\label{solution3}
\end{eqnarray}
the solution was obtained~\cite{Dzhunushaliev:2009va,Bogdanos:2006qw,Liu:2012gv}
\begin{eqnarray}
\phi(y)&=&\phi_{0}\tanh(ay),
\end{eqnarray}
where $\gamma=2(\frac{1}{\xi}-6)$ and $\phi_{0}=a^{-1}\phi(0)=\sqrt{\frac{3(1-6\xi)}{\xi(1-2\xi)}}$. Note that the parameter $\xi$ satisfies $0<\xi<1/6$, which leads to $\gamma>0$. And the scalar curvature $R$ approaches a negative value when $y\rightarrow\infty$. It means that the bulk is asymptotically AdS.

\begin{figure}[htb]
\includegraphics[width=8cm,height=6.5cm]{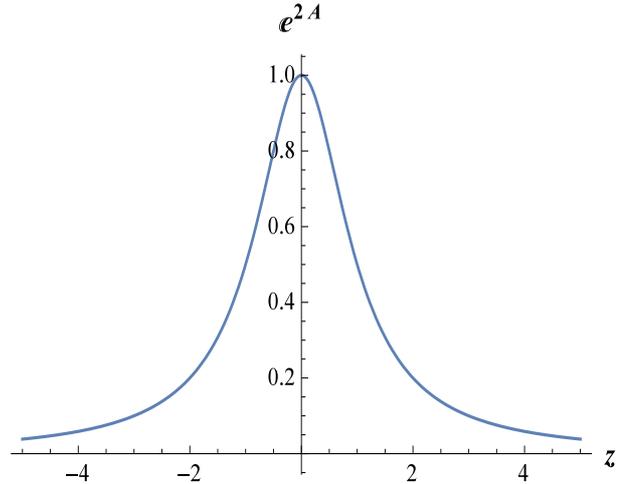}
\caption{ The shape of
the warp factor $\text{e}^{2A(z)}$ in the coordinate $z$. The parameters are set to $a=1$ and $b=1$.}
\label{f1}
\end{figure}

It is obvious that the warp factors of the two solutions have the same form:
\begin{eqnarray}
\text{e}^{2A(y)}=\cosh(a y)^{-2b}. \label{warp factor 1}
\end{eqnarray}
Here $a$ is an arbitrary constant parameter and $b$ a positive real constant. For the first brane model it means $\bar{b}=b$ according to Eq. (\ref{solution2}) and for the second one it means $\gamma=b$ according to Eq.~(\ref{solution3}). It is easy to check that it is the solution to the Einstein equations (\ref{Einstein and scalar equation}) and (\ref{Einstein and scalar equation2}) by bringing the solution back to Eqs.~(\ref{Einstein and scalar equation}) and (\ref{Einstein and scalar equation2}). With the coordinate transformation (\ref{coordinate transformation}), we can get the conformal coordinate
\begin{eqnarray}
z(y)=&-&{i}\frac{\sqrt{\pi}\Gamma(\frac{1+b}{2})}{2|a|\Gamma(1+\frac{b}{2})}\nonumber\\
     &+&{i}\text{sign}(a y)\frac{[\cosh(a y)]^{1+b}}
    {a(1+b)}F,
\label{zy}
\end{eqnarray}
where $F$ is the hypergeometric function
\begin{eqnarray}
F={}_{2}F_{1}\left[\frac{1}{2},\frac{1+b}{2},\frac{3+b}{2},\cosh^{2}(a y)\right].
\end{eqnarray}
For simplicity, we just chose the simple case of $b=1$. In this case, we have
\begin{eqnarray}
z=\int_0^y \cosh(a \bar{y}) d\bar{y}=\frac{1}{a}\sinh(ay),
\end{eqnarray}
and
\begin{eqnarray}
A(z)=-\ln(\cosh(\text{arcsinh}(az)))=-\frac{1}{2}\ln(1+a^2z^2).
\end{eqnarray}
Figure~\ref{f1} shows the shape of the warp factor $\text{e}^{2A(z)}$ in the coordinate $z$.
As what we have discussed in the previous section, the role of $K(z)$ is similar to the superpotential $W(\phi)$ in thick brane models. Thus for different forms of $K(z)$, there exist different zero mode solutions and configurations of $f(\phi)$. For example, next we will focus on three kinds of $K(z)$ and investigate the localization of the zero mode (\ref{zeromodeforz2}).

\subsection{$K(z)=k\frac{C'}{C}$}
\begin{figure}[htb]
\includegraphics[width=8cm,height=6.5cm]{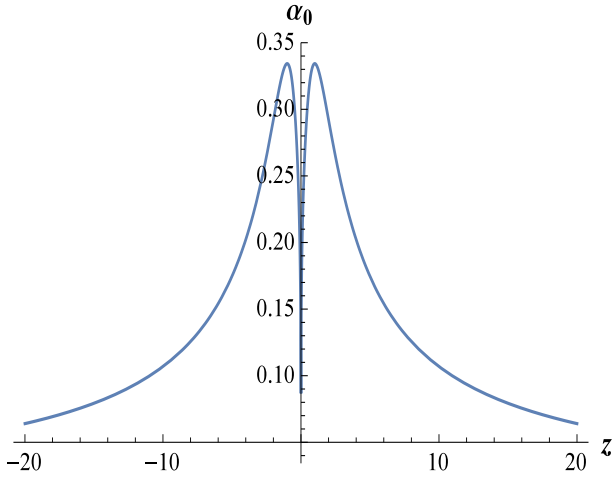}
\caption{ The shape of
the Elko zero mode $\alpha_{0}(z)$ in (\ref{zeromodeforz3})
in the thick brane models with $A(z(y))=-b\ln\cosh(ay)$. The parameters are set to $b=1$, $a=1$ and $k=\frac{1}{2}$.}
\label{VandX}
\end{figure}

Firstly, we consider $K(z)=k\frac{C'}{C}$ with $k\geq0$.
It is easy to get
\begin{eqnarray}
C(z)=-\frac{1-k}{z+\mathcal{C}_2}.
\end{eqnarray}
Here $\mathcal{C}_2$ is a constant and we can always let $\mathcal{C}_2=0$. The zero mode is rewritten as
\begin{eqnarray}
 \alpha_{0}(z)&\varpropto& \text{e}^{D(z)}\nonumber\\
 &=&{\exp}\left[\frac{1}{2}\int^{z}_{0}\left(\frac{A'^2}{C}-k\frac{C'}{C}\right)d\bar{z}\right]\nonumber\\
 &=& -(|C|^{-1})^{\frac{k}{2}}\left[\frac{1}{2}\int^{z}_{0}A'^2(\bar{z})\frac{\bar{z}}{1-k}d\bar{z}\right]\nonumber\\
  &\varpropto&
           \frac{ |z|^{\frac{k}{2}} ~\text{e}^{-\frac{1}{4(1-k)({1+a^2 z^2})}}  }
                { (1+a^2z^2)^{\frac{1}{4(1-k)}} }~
          \label{zeromodeforz3}
\end{eqnarray}
with the condition $1-k>0$. In this case, the orthonormality condition (\ref{orthonormality relation for zero mode}) reads
\begin{eqnarray}
\int \alpha_0^2dz
   \varpropto
    \int   \frac{ |z|^{k} ~\text{e}^{-\frac{1}{2(1-k)({1+a^2 z^2})}} }
                { (1+a^2z^2)^{\frac{1}{2(1-k)}} }~
          dz<\infty.\label{orthonormality relation for z2}
\end{eqnarray}
It is clear that the express $\frac{1}{1+a^2 z^2}$ tends to zero and $|z|^{k}(1+a^2z^2)^{-\frac{1}{2(1-k)}}\rightarrow |z|^{k}(az)^{-\frac{1}{(1-k)}}$ as $z\rightarrow\infty$. Since the integral $\int (1+z)^n dz$ will be convergent only for $n<-1$, the orthonormality condition requires  $k-\frac{1}{1-k}<-1$, i.e., $0<k<1$. The shape of the zero mode $\alpha_{0}(z)$ in (\ref{zeromodeforz3}) is shown in Fig.~\ref{VandX}.
It is obvious that this kind of zero mode can be localized on the branes under the condition $0<k<1$. However, this kind of zero mode is uncommon. It is clear that there exists a split at the point $z=0$, which divides the zero mode into two halves. This split comes from the absolute value function $|C|^{-1}$ in Eq.~(\ref{zeromodeforz3})
and it also exists in the non-minimal coupling function $f(\phi)$ (see Eq. (\ref{ffunction})), which will provide an interesting and uncommon coupling $f(\phi)$. In this case, the non-minimal coupling function $f(\phi)$ reads
\begin{eqnarray}
f(\phi(z))&=&\text{e}^{\int^{z}_{0} B(\bar{z}) d\bar{z}}\nonumber\\
          &=&{\exp}\left[\int^{z}_{0}\left( -3A'+\frac{A'^2}{C}-(2-k)\frac{C'}{C} \right) d\bar{z}\right]\nonumber\\
          &=&{\exp}\left[-3A-(2-k)\ln |C|+\int^{z}_{0}\frac{A'^2}{C}d\bar{z}\right]\nonumber\\
          &=&\left(\frac{1}{1-k}\right)^{2-k}\text{e}^{-3A}|z|^{2-k}\nonumber\\
          &&\times{\exp}\left[-\frac{1}{(1-k)}\int^{z}_{0}A'^2\bar{z}d\bar{z}\right].\label{ffunction}
\end{eqnarray}
Thus, the function $f(\phi)$ for the canonical-scalar-generated brane model reads
\begin{eqnarray}
f(\phi)&=&\frac{|\Phi|^{2-k}   ~  \text{e}^{\frac{\Phi^2}{2(1-k)({1+\Phi^2})}} }
                  { \left({a(1-k)}\right)^{2-k} }
             (1+\Phi^2)^{\frac{2-3k}{2(1-k)}},
\end{eqnarray}
where $\Phi=\sinh\left(\frac{2}{3}\text{arctanh}\tan\left(\frac{\phi}{\phi_0}\right)\right)$.
The function $f(\phi)$ for the second brane model reads
\begin{eqnarray}
f(\phi)
          &=& \frac{ \tilde{|\phi|}^{2-k}  ~  \text{e}^{\frac{\tilde{\phi}^2}{2(1-k)}}  }
                   { \left({a(1-k)}\right)^{2-k} }
             (1-\tilde{\phi}^2)^{-\frac{4+k(k-6)}{2(1-k)}}
             ,
\end{eqnarray}
where $\tilde{\phi}=\frac{\phi}{\phi_0}$. We can find that both forms of $f(\phi)$ are complex.
They are different because of different solutions of the scalar $\phi$. This result shows that the non-minimal coupling function $f(\phi)$ will be different for different kinds of brane worlds, even though the zero mode has the same form.


\subsection{$K(z)=k z$}
\begin{figure}[htb]
\includegraphics[width=8cm,height=6.5cm]{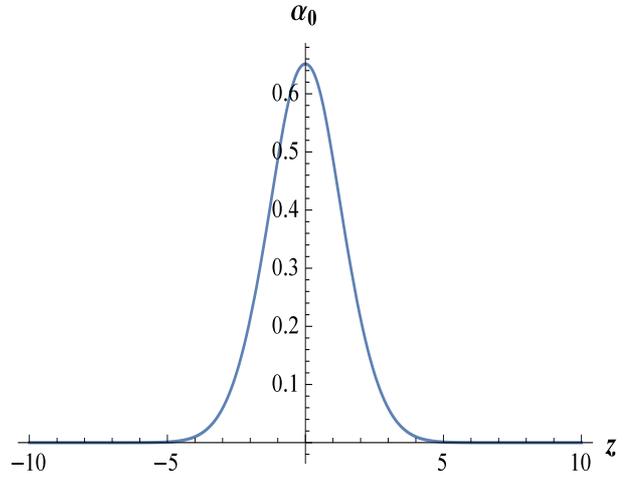}
\caption{ The shape of
the Elko zero mode (\ref{zeromodeforz4})
in the thick brane models. The parameters are set to $a=k=1$.}
\label{Z2}
\end{figure}

Next, we consider the case of $K(z)=kz$ with $k>0$. The form of $C(z)$ reads
\begin{eqnarray}
C(z)=-\frac{\text{e}^{\frac{kz^2}{2}}\sqrt{\frac{2k}{\pi}}}{\text{erf}
\left(\sqrt{\frac{k}{2}}z\right)}.
\end{eqnarray}
Here $\text{erf}\left(\sqrt{\frac{k}{2}}z\right)$ is the imaginary error function. Thus the zero mode is changed to be
\begin{eqnarray}
 \alpha_{0}(z)
 &\varpropto&{\exp}\left[\frac{1}{2}\int^{z}_{0}\frac{A'^2}{C}d\bar{z}\right]{\exp}\left[-\frac{1}{2}\int^{z}_{0}Kd\bar{z}\right]\nonumber\\
 &=&{\exp}\bigg[-\frac{1}{2}
       \sqrt{\frac{\pi}{2k}} \int^{z}_{0}\text{erf}
       \left(\sqrt{\frac{k}{2}}\bar{z}\right)
       \frac{a^4\bar{z}^2~\text{e}^{{-k \bar{z}^{2}}/{2}}}
            {(1+a^2\bar{z}^2)^2}
       d\bar{z}\nonumber\\
                &&~~~~~~-\frac{k}{4}z^{2}\bigg]. \label{zeromodeforz4}
\end{eqnarray}
It is a fact that the integral $\int^{z}_{0}\text{erf}
       \left(\sqrt{\frac{k}{2}}\bar{z}\right)
       \frac{a^4\bar{z}^2~\text{e}^{{-k \bar{z}^{2}}/{2}}}
            {(1+a^2\bar{z}^2)^2}
       d\bar{z}$ is a finite function of $z$.
 So we have $\alpha_0(z)\varpropto{\exp}\left[-\frac{k}{4}z^{2}\right]$ as $|z| \to \infty$. The zero mode (\ref{zeromodeforz4}) is plotted in Fig.~(\ref{Z2}) with $a=k=1$.
In this case, the orthonormality condition $\int \alpha_0^2dz <\infty  $
is satisfied. Thus this kind of zero mode can be localized on the thick branes. The function $f(\phi)$ can be solved as
\begin{eqnarray}
f(\phi) &=&\frac{\pi}{2k}{\exp}\left[ - \sqrt{\frac{\pi a^2}{2k}}
            \int^{\frac{\Phi}{a}}_{0}
           \text{erf}\left(\sqrt{\frac{k}{2}}\frac{\bar{\Phi}}{a}\right)
           \frac{\bar{\Phi}^2~\text{e}^{\frac{-k \bar{\Phi}^{2}}{2a^2}}}{(1+\bar{\Phi}^2)^2}
           d\bar{\Phi}\right]\nonumber\\
         &&\times(1+\Phi^2)^{\frac{3}{2}}\text{erf}^{\,2}\left(\sqrt{\frac{k}{2}}\frac{\Phi}{a}\right)\text{e}^{\frac{k \Phi^{2}}{2a^2}},
\end{eqnarray}
for the first brane model and
\begin{eqnarray}
f(\phi)
   &=&\frac{\pi}{2k}{\exp}\left[
           -\sqrt{\frac{\pi}{2k}} \int^{\frac{\tilde{\phi}}{a}}_{0}
           \text{erf}\left(\mathcal{J} (\bar{\phi})\right)
           a\bar{{\phi}}^2
           (1-\bar{{\phi}}^2)
           \text{e}^{-\mathcal{J} (\bar{\phi})}
           d\bar{{\phi}}\right]\nonumber\\
   &&\times(1-\tilde{\phi}^2)^{-\frac{3}{2}}~
          \text{erf}^{\,2} \left(\mathcal{J} (\tilde{\phi})\right)
          \text{e}^{\mathcal{J}^2 (\tilde{\phi})},
\end{eqnarray}
for the second one,
where $\Phi=\sinh\left(\frac{2}{3}\text{arctanh}\tan\left(\frac{\phi}{\phi_0}\right)\right)$, $\mathcal{J} (\tilde{\phi}) =\frac{\sqrt{k}\tilde{\phi}}{a\sqrt{2(1-\tilde{\phi}^2)}}$, and $\tilde{\phi}=\frac{\phi}{\phi_0}$. It is obvious that the forms of the function $f(\phi)$
are more complex than the ones in the case of $K(z)=k\frac{C'}{C}$ so that they can not be written as an elementary function. However, the absolute value function $|C|^{-1}$ in the zero mode and the function $f(\phi)$ disappears in this case, which means that it can be eliminated by adopting an appropriate form of $K(z)$. At the same time, it is worth pointing out that the parameter $k$ is related with the coupling constant, and its range is larger than that in the previous case in this paper. For this case, the zero mode can be localized on branes for any positive $k$.


\subsection{$K(z)=k\tanh(kz)$}
\begin{figure}[htb]
\includegraphics[width=8cm,height=6.5cm]{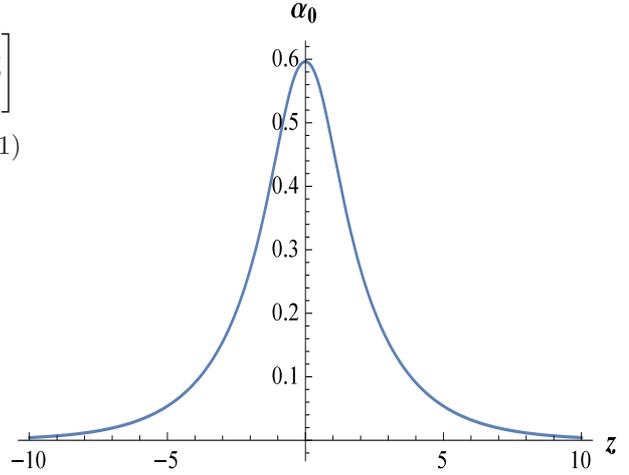}
\caption{ The shape of
the Elko zero mode $\alpha_{0}(z)$ (\ref{zeromodeforz5})
in the thick brane model. The parameters are set to $a=1$ and $k=1$.}
\label{Z3}
\end{figure}

Finally, we take $K(z)=k\tanh(kz)$ with $k>0$. The form of $C(z)$ can be solved as
\begin{eqnarray}
C(z)=-k\coth(kz)
\end{eqnarray}
and the zero mode reads
\begin{eqnarray}
 \alpha_{0}(z)&\varpropto& 
\text{sech}^{\frac{1}{2}}(kz) ~{\exp}\left[-\int^{z}_{0}\frac{a^4\bar{z}^2 \tanh(k\bar{z})}{{2k}(1+a^2\bar{z}^2)^2}d\bar{z}\right].\label{zeromodeforz5}
\end{eqnarray}
It is easy to see that the function $\int^{z}_{0}\frac{a^4\bar{z}^2}{(1+a^2\bar{z}^2)^2}\tanh(k\bar{z})d\bar{z}$ trends to $\mp\frac{1}{z}$ as $z\rightarrow\pm\infty$. So the second factor in \eqref{zeromodeforz5} has only a constant contribution at boundaries of the extra dimension, but the first one has an important role, i.e., $\alpha_0(z)\varpropto \text{e}^{-\frac{1}{2}k|z|}$ as $z\rightarrow\pm\infty$. The zero mode (\ref{zeromodeforz5}) is shown in Fig.~(\ref{Z3}) with $a=1$ and $k=1$.
It is clear that the orthonormality condition is also satisfied, and so the zero mode can be localized on the branes for any positive $k$.
The function $f(\phi)$ for the two brane models are
\begin{eqnarray}
f(\phi)
         &=&\frac{1}{k^2}(1+\Phi^2)^{\frac{3}{2}}
            \tanh\left(\frac{k}{a}\Phi\right)
            \text{sinh}\left(\frac{k}{a}\Phi\right)\nonumber\\
         &&\times{\exp}\left[-\frac{a}{k}\int^{\frac{\Phi}{a}}_{0}
                         \frac{\bar{\Phi}^2 \tanh(\frac{k}{a}\bar{\Phi})}
                              {(1+\bar{\Phi}^2)^2}
                          d\bar{\Phi}\right]
\end{eqnarray}
and
\begin{eqnarray}
 f(\phi) &=&\frac{\tanh\left(\mathcal{K}(\tilde{\phi})\right)
                   \text{sinh}\left(\mathcal{K}(\tilde{\phi})\right)}
                 {k^2 (1-\tilde{\phi}^2)^{\frac{3}{2}}}\nonumber\\
          &&\times{\exp}\left[-\frac{a}{k}
          \int^{\frac{\tilde{\phi}}{a}}_{0}\bar{{\phi}}^2(1-\bar{{\phi}}^2)
            \tanh\left(\mathcal{K}(\bar{{\phi}})\right)d\bar{{\phi}}\right],~~
\end{eqnarray}
respectively,
wehre $\Phi=\sinh\left(\frac{2}{3}\text{arctanh}\tan\left(\frac{\phi}{\phi_0}\right)\right)$, $\mathcal{K}(\tilde{\phi})=\left(\frac{k\tilde{\phi}}{a\sqrt{1-\tilde{\phi}^2}}\right)$, and $\tilde{\phi}=\frac{\phi}{\phi_0}$.

\section{Conclusion and discussion}
\label{section5}

In this paper, we reconsidered the localization of a five-dimensional Elko spinor on thick branes. In our previous work \cite{Liu:2011nb}, the Yukawa-type coupling between the Elko spinor and the background scalar field has been introduced to realize the localization of the Elko zero mode. It was shown that the localized Elko zero mode can just be obtained for some kinds of thick branes, only if the coupling constant is taken as some particular expression determined by the parameters in the brane models. Here, we introduced new localization mechanism, i.e., the non-minimal coupling $f(\phi)\mathfrak{L}_{Elko}$ between the five-dimensional Elko spinor and the background scalar field.
The Schr\"{o}dinger like equation for the Elko zero mode was derived. The result shows that the effective potential depends on the form of $f(\phi)$ and the warp factor $\text{e}^{2A}$. For convenience, we introduced a series of new functions, especially the function $K(z)$, to obtain the Elko zero mode. We gave the general expressions of the zero mode $\alpha_0$ and the function $f(\phi)$ by using $K(z)$. It was found that the role of $K(z)$ is similar to the superpotential $W(\phi)$ in thick brane models. The zero mode $\alpha_0$ and the scalar field function $f(\phi(z))$ can be obtained by a given $K(z)$. Thus an appropriate form of $K(z)$ is needed to confine the Elko zero mode on branes and different forms of $K(z)$ will lead to different zero mode solutions and configurations of $f(\phi)$.

Next we reviewed two kinds of Minkowski thick brane models. One is for a standard canonical scalar field and the other is for a non-minimally coupled scalar field. In Ref. \cite{Liu:2011nb} we had investigated the localization of the Elko spinor with Yukawa-type coupling on these thick branes. For these branes, the warp factor has the same form while the solutions of the background scalar field are different. In order to illustrate the effect of $K(z)$ on the zero mode $\alpha_0$ and the function $f(\phi)$, we took three different forms of $K(z)$ as examples. Firstly, we focused on the case of $K(z)=k\frac{C'}{C}$ with $k\geq0$. Here the parameter $k$ is related with the coupling constant. It was found that the Elko zero mode can be localized on the branes with the limit of $0<k<1$. It is very interesting that a split at the point $z=0$ divides the Elko zero mode into two halves, which comes from the absolute value in the form of the zero mode as well as the function $f(\phi)$. The forms of $f(\phi)$ are very complex and different for the two different thick brane models. It is clear that there will exist different non-minimal couplings for different brane models, even though the zero modes have the same form.

Secondly, we took $K(z)=kz$ with $k>0$. In this case, the zero mode is so complex that it is hard to be written as an explicit function. Fortunately, the complex part of the zero mode just acts an unimportant role and the zero mode has the same behaviors as ${\exp}\left(-\frac{k}{4}z^{2}\right)$ at the boundaries of the extra dimension. Thus the zero mode can also be localized in this case. The absolute value in the zero mode and the function $f(\phi)$ disappears and the range of the parameter $k$ is larger than that in the first case. The result in this case shows that different choices of $K(z)$ will lead to different configurations of solutions.

Finally, we adopted $K(z)=k\tanh(kz)$ with $k>0$. In this case, the zero mode also has a complex form and can not be written as an elementary function. As the second case, the complex part does not affect the asymptotic behavior of the Elko zero mode and the zero mode has the same property of $\text{sech}^{\frac{1}{2}}(kz)$, i.e., $\alpha_0(z) \propto \text{sech}^{\frac{1}{2}}(kz)$ at the boundaries of the extra dimension. Such zero mode can be localized on both thick branes with any positive $k$. The forms of the function $f(\phi)$ are still very complex and they are inequable for different kinds of thick branes.

In this paper, we considered two kinds of Minkowski thick branes and obtained three different expressions of bounded Elko zero modes. The result shows that the non-minimal coupling can provide us more possibilities of localizing the Elko zero mode. In addition, it may give us new way to solve the localization problem of Elko massive KK modes, which will be investigated in our further work.

\section*{Acknowledgments}

This work was supported by the National Natural Science Foundation of China (Grant No. 11522541, No. 11705106, and No. 11305095), and the Fundamental Research Funds for the Central Universities (Grant No. lzujbky-2016-k04).


\begin{thebibliography}{10}
\providecommand{\url}[1]{\texttt{#1}}
\providecommand{\urlprefix}{URL }
\providecommand{\eprint}[2][]{\url{#2}}

\bibitem{ArkaniHamed:1998rs}
  N.~Arkani-Hamed, S.~Dimopoulos, and G.~R.~Dvali,
  Phys.\ Lett.\ B {\bf 429} (1998) 263,
  arXiv:hep-ph/9803315.

\bibitem{Randall:1999ee}
  L.~Randall and R.~Sundrum,
  Phys.\ Rev.\ Lett.\  {\bf 83} (1999) 3370,
  arXiv:hep-ph/9905221.

\bibitem{Randall:1999vf}
  L.~Randall and R.~Sundrum,
  Phys.\ Rev.\ Lett.\  {\bf 83} (1999) 4690,
  arXiv:hep-th/9906064.

\bibitem{Antoniadis:1998ig}
  I.~Antoniadis, N.~Arkani-Hamed, S.~Dimopoulos, and G.~R.~Dvali,
  Phys.\ Lett.\ B {\bf 436} (1998) 257,
  arXiv:hep-ph/9804398.

\bibitem{Das:2007qn}
  S.~Das, D.~Maity, and S.~SenGupta,
  JHEP {\bf 0805} (2008) 042,
  arXiv:0711.1744 [hep-th].

\bibitem{Yang:2012dd}
  K.~Yang, Y.~X.~Liu, Y.~Zhong, X.~L.~Du, and S.~W.~Wei,
  Phys.\ Rev.\ D {\bf 86} (2012) 127502,
  arXiv:1212.2735 [hep-th].


\bibitem{Guo:2015wua}
  B.~Guo, Y.~X.~Liu, K.~Yang, and X.~H.~Meng,
  arXiv:1501.02674 [hep-th].

\bibitem{ArkaniHamed:2000eg}
  N.~Arkani-Hamed, S.~Dimopoulos, N.~Kaloper, and R.~Sundrum,
  Phys.\ Lett.\ B {\bf 480} (2000) 193,
  arXiv:hep-th/0001197.


\bibitem{Kim:2000mc}
  J.~E.~Kim, B.~Kyae, and H.~M.~Lee,
  Phys.\ Rev.\ Lett.\  {\bf 86} (2001) 4223,
  arXiv:hep-th/0011118.

\bibitem{Starkman:2000dy}
  G.~D.~Starkman, D.~Stojkovic, and M.~Trodden,
  Phys.\ Rev.\ D {\bf 63} (2001) 103511,
  arXiv:hep-th/0012226.

\bibitem{Starkman:2001xu}
  G.~D.~Starkman, D.~Stojkovic, and M.~Trodden,
  Phys.\ Rev.\ Lett.\  {\bf 87} (2001) 231303,
  arXiv:hep-th/0106143.

\bibitem{Kehagias:2004fb}
  A.~Kehagias,
  Phys.\ Lett.\ B {\bf 600} (2004) 133,
  arXiv:hep-th/0406025.

\bibitem{George:2008vu}
  D.~P.~George, M.~Trodden, and R.~R.~Volkas,
  JHEP {\bf 0902} (2009) 035,
  arXiv:0810.3746 [hep-ph].

\bibitem{Dey:2009xu}
  P.~Dey, B.~Mukhopadhyaya, and S.~SenGupta,
  Phys.\ Rev.\ D {\bf 80} (2009) 055029,
  arXiv:0904.1970 [hep-ph].

\bibitem{Haghani:2012zq}
  Z.~Haghani, H.~R.~Sepangi, and S.~Shahidi,
  JCAP {\bf 1202} (2012) 031,
  arXiv:1201.6448 [gr-qc].


\bibitem{ArkaniHamed:1998nn}
  N.~Arkani-Hamed, S.~Dimopoulos, and G.~R.~Dvali,
  Phys.\ Rev.\ D {\bf 59} (1999) 086004£¬
  arXiv:hep-ph/9807344.

\bibitem{Sahni:2002dx}
  V.~Sahni and Y.~Shtanov,
  JCAP {\bf 0311} (2003) 014,
  arXiv:astro-ph/0202346.

\bibitem{Cembranos:2003mr}
  J.~A.~R.~Cembranos, A.~Dobado, and A.~L.~Maroto,
  Phys.\ Rev.\ Lett.\  {\bf 90} (2003) 241301,
  arXiv:hep-ph/0302041.

\bibitem{Nihei:2004xv}
  T.~Nihei, N.~Okada, and O.~Seto,
  Phys.\ Rev.\ D {\bf 71} (2005) 063535,
  arXiv:hep-ph/0409219.

\bibitem{Gremm:1999pj}
  M.~Gremm,
  Phys.\ Lett.\ B {\bf 478} (2000) 434,
  arXiv:hep-th/9912060.

\bibitem{DeWolfe:1999cp}
  O.~DeWolfe, D.~Z.~Freedman, S.~S.~Gubser, and A.~Karch,
  Phys.\ Rev.\ D {\bf 62} (2000) 046008,
  arXiv:hep-th/9909134.


\bibitem{Kobayashi:2001jd}
  S.~Kobayashi, K.~Koyama, and J.~Soda,
  Phys.\ Rev.\ D {\bf 65} (2002) 064014,
  arXiv:hep-th/0107025.

\bibitem{Bazeia:2002sd}
  D.~Bazeia, L.~Losano, and C.~Wotzasek,
  Phys.\ Rev.\ D {\bf 66} (2002) 105025,
  arXiv:hep-th/0206031.

\bibitem{Bazeia:2004dh}
  D.~Bazeia and A.~R.~Gomes,
  JHEP {\bf 0405} (2004) 012,
  arXiv:hep-th/0403141.

\bibitem{Afonso:2006gi}
  V.~I.~Afonso, D.~Bazeia, and L.~Losano,
  Phys.\ Lett.\ B {\bf 634} (2006) 526,
  arXiv:hep-th/0601069.

\bibitem{Bazeia:2006ef}
  D.~Bazeia, F.~A.~Brito, and L.~Losano,
  JHEP {\bf 0611} (2006) 064,
  arXiv:hep-th/0610233.


\bibitem{Bogdanos:2006qw}
  C.~Bogdanos, A.~Dimitriadis, and K.~Tamvakis,
  Phys.\ Rev.\ D {\bf 74} (2006) 045003,
  arXiv:hep-th/0604182.

\bibitem{Dzhunushaliev:2009va}
  V.~Dzhunushaliev, V.~Folomeev, and M.~Minamitsuji,
  Rept.\ Prog.\ Phys.\  {\bf 73} (2010) 066901,
  arXiv:0904.1775 [gr-qc].

\bibitem{Liu:2009ega}
  Y.~X.~Liu, Y.~Zhong, and K.~Yang,
  Europhys.\ Lett.\  {\bf 90} (2010) 51001,
  arXiv:0907.1952 [hep-th].

\bibitem{Liu:2011wi}
  Y.~X.~Liu, Y.~Zhong, Z.~H.~Zhao, and H.~T.~Li,
  JHEP {\bf 1106} (2011) 135,
  arXiv:1104.3188 [hep-th].

\bibitem{Guo:2011wr}
  H.~Guo, Y.~X.~Liu, Z.~H.~Zhao, and F.~W.~Chen,
  Phys.\ Rev.\ D {\bf 85} (2012) 124033,
  arXiv:1106.5216 [hep-th].

\bibitem{Liu:2012rc}
  Y.~X.~Liu, K.~Yang, H.~Guo, and Y.~Zhong,
  Phys.\ Rev.\ D {\bf 85} (2012) 124053,
  arXiv:1203.2349 [hep-th].


\bibitem{Liu:2012gv}
  Y.~X.~Liu, F.~W.~Chen, Heng-Guo, and X.~N.~Zhou,
  JHEP {\bf 1205} (2012) 108,
  arXiv:1205.0210 [hep-th].


\bibitem{Bazeia:2012br}
  D.~Bazeia, F.~A.~Brito, and F.~G.~Costa,
  Phys.\ Rev.\ D {\bf 87} (2013) no.6,  065007,
  arXiv:1210.6318 [hep-th].

\bibitem{German:2013sk}
  G.~German, A.~Herrera--Aguilar, D.~Malagon--Morejon, I.~Quiros, and R.~da Rocha,
  Phys.\ Rev.\ D {\bf 89} (2014) no.2,  026004,
  arXiv:1301.6444 [hep-th].


\bibitem{Dutra:2014xla}
  A.~de Souza Dutra, G.~P.~de Brito, and J.~M.~Hoff da Silva,
  Phys.\ Rev.\ D {\bf 91} (2015) no.8,  086016,
  doi:10.1103/PhysRevD.91.086016,
  arXiv:1412.5543 [hep-th].

\bibitem{Guo:2014bxa}
  B.~Guo, Y.~X.~Liu, and K.~Yang,
  Eur.\ Phys.\ J.\ C {\bf 75} (2015) no.2,  63,
  arXiv:1405.0074 [hep-th].


\bibitem{Geng:2015kvs}
  W.~J.~Geng and H.~Lu,
  Phys.\ Rev.\ D {\bf 93} (2016) no.4,  044035,
  arXiv:1511.03681 [hep-th].

\bibitem{BarbosaCendejas:2005kn}
  N.~Barbosa-Cendejas and A.~Herrera-Aguilar,
  JHEP {\bf 0510} (2005) 101,
  arXiv:hep-th/0511050.

\bibitem{HerreraAguilar:2010kt}
  A.~Herrera-Aguilar, D.~Malagon-Morejon, and R.~R.~Mora-Luna,
  JHEP {\bf 1011} (2010) 015,
  arXiv:1009.1684 [hep-th].

\bibitem{Zhong:2015pta}
  Y.~Zhong and Y.~X.~Liu,
  Eur.\ Phys.\ J.\ C {\bf 76} (2016) no.6,  321,
  arXiv:1507.00630 [hep-th].

\bibitem{Liu:2017}
  Y.~X.~Liu,
  \emph{Introduction to Extra Dimensions and Thick Braneworlds},
  arXiv:1707.08541 [hep-th].


\bibitem{Hoyle:2000cv}
  C.~D.~Hoyle, U.~Schmidt, B.~R.~Heckel, E.~G.~Adelberger, J.~H.~Gundlach, D.~J.~Kapner, and H.~E.~Swanson,
  Phys.\ Rev.\ Lett.\  {\bf 86} (2001) 1418,
  arXiv:hep-ph/0011014.


\bibitem{Hung:2003cj}
  P.~Q.~Hung and N.~K.~Tran,
  Phys.\ Rev.\ D {\bf 69} (2004) 064003,
  arXiv:hep-ph/0309115.

\bibitem{Adelberger:2006dh}
  E.~G.~Adelberger, B.~R.~Heckel, S.~A.~Hoedl, C.~D.~Hoyle, D.~J.~Kapner, and A.~Upadhye,
  Phys.\ Rev.\ Lett.\  {\bf 98} (2007) 131104,
  arXiv:hep-ph/0611223.

\bibitem{Guo:2011qt}
  H.~Guo, A.~Herrera-Aguilar, Y.~X.~Liu, D.~Malagon-Morejon, and R.~R.~Mora-Luna,
  Phys.\ Rev.\ D {\bf 87} (2013) no.9,  095011,
  arXiv:1103.2430 [hep-th].

\bibitem{Bajc:1999mh}
  B.~Bajc and G.~Gabadadze,
  Phys.\ Lett.\ B {\bf 474} (2000) 282,
  arXiv:hep-th/9912232.

\bibitem{Oda:2000zc}
  I.~Oda,
  Phys.\ Lett.\ B {\bf 496} (2000) 113,
  arXiv:hep-th/0006203.

\bibitem{Liu:2007ku}
  Y.~X.~Liu, X.~H.~Zhang, L.~D.~Zhang, and Y.~S.~Duan,
  JHEP {\bf 0802} (2008) 067,
  arXiv:0708.0065 [hep-th].

\bibitem{Liu:2008wd}
  Y.~X.~Liu, L.~D.~Zhang, S.~W.~Wei, and Y.~S.~Duan,
  JHEP {\bf 0808} (2008) 041,
  arXiv:0803.0098 [hep-th].

\bibitem{Liu:2009uca}
  Y.~X.~Liu, H.~Guo, C.~E.~Fu, and J.~R.~Ren,
  JHEP {\bf 1002} (2010) 080,
  arXiv:0907.4424 [hep-th].

\bibitem{Guerrero:2009ac}
  R.~Guerrero, A.~Melfo, N.~Pantoja, and R.~O.~Rodriguez,
  Phys.\ Rev.\ D {\bf 81} (2010) 086004,
  arXiv:0912.0463 [hep-th].

\bibitem{Zhao:2010mk}
  Z.~H.~Zhao, Y.~X.~Liu, H.~T.~Li, and Y.~Q.~Wang,
  Phys.\ Rev.\ D {\bf 82} (2010) 084030,
  arXiv:1004.2181 [hep-th].

\bibitem{Chumbes:2010xg}
  A.~E.~R.~Chumbes, A.~E.~O.~Vasquez, and M.~B.~Hott,
  Phys.\ Rev.\ D {\bf 83} (2011) 105010,
  arXiv:1012.1480 [hep-th].

\bibitem{Chumbes:2011zt}
  A.~E.~R.~Chumbes, J.~M.~Hoff da Silva, and M.~B.~Hott,
  Phys.\ Rev.\ D {\bf 85} (2012) 085003,
  arXiv:1108.3821 [hep-th].

\bibitem{Germani:2011cv}
  C.~Germani,
  Phys.\ Rev.\ D {\bf 85} (2012) 055025,
  arXiv:1109.3718 [hep-ph].

\bibitem{Cruz:2012kd}
  W.~T.~Cruz, A.~R.~P.~Lima, and C.~A.~S.~Almeida,
  Phys.\ Rev.\ D {\bf 87} (2013) no.4,  045018,
  arXiv:1211.7355 [hep-th].

\bibitem{Fu:2013ita}
  C.~E.~Fu, Y.~X.~Liu, H.~Guo, F.~W.~Chen, and S.~L.~Zhang,
  Phys.\ Lett.\ B {\bf 735} (2014) 7,
  arXiv:1312.2647 [hep-th].

\bibitem{Liu:2013kxz}
  Y.~X.~Liu, Z.~G.~Xu, F.~W.~Chen, and S.~W.~Wei,
  Phys.\ Rev.\ D {\bf 89} (2014) no.8,  086001,
  arXiv:1312.4145 [hep-th].

\bibitem{Zhao:2014gka}
  Z.~H.~Zhao, Y.~X.~Liu, and Y.~Zhong,
  Phys.\ Rev.\ D {\bf 90} (2014) no.4,  045031,
  arXiv:1402.6480 [hep-th].

\bibitem{Alencar:2014moa}
  G.~Alencar, R.~R.~Landim, M.~O.~Tahim, and R.~N.~Costa Filho,
  Phys.\ Lett.\ B {\bf 739} (2014) 125,
  arXiv:1409.4396 [hep-th].

\bibitem{Vaquera-Araujo:2014tia}
  C.~A.~Vaquera-Araujo and O.~Corradini,
  Eur.\ Phys.\ J.\ C {\bf 75} (2015) no.2,  48,
  arXiv:1406.2892 [hep-th].

\bibitem{Fu:2015cfa}
  C.~E.~Fu, Y.~X.~Liu, H.~Guo, and S.~L.~Zhang,
  Phys.\ Rev.\ D {\bf 93} (2016) no.6,  064007,
  arXiv:1502.05456 [hep-th].

\bibitem{Fu:2016vaj}
  C.~E.~Fu, Y.~Zhong, Q.~Y.~Xie, and Y.~X.~Liu,
  Phys.\ Lett.\ B {\bf 757} (2016) 180,
  arXiv:1601.07118 [hep-th].

\bibitem{Zhang:2016ksq}
  Y.~P.~Zhang, Y.~Z.~Du, W.~D.~Guo, and Y.~X.~Liu,
  Phys.\ Rev.\ D {\bf 93} (2016) no.6,  065042,
  arXiv:1601.05852 [hep-th].

\bibitem{Li:2017dkw}
  Y.~Y.~Li, Y.~P.~Zhang, W.~D.~Guo, and Y.~X.~Liu,
  Phys.\ Rev.\ D {\bf 95} (2017) no. 11, 115003,
  arXiv:1701.02429 [hep-th].



\bibitem{Ahluwalia:2004sz}
  D.~V.~Ahluwalia and D.~Grumiller,
  Phys.\ Rev.\ D {\bf 72} (2005) 067701,
  arXiv:hep-th/0410192.

\bibitem{Ahluwalia:2004ab}
  D.~V.~Ahluwalia and D.~Grumiller,
  JCAP {\bf 0507} (2005) 012,
  arXiv:hep-th/0412080.

\bibitem{Ahluwalia:2008xi}
  D.~V.~Ahluwalia, C.~Y.~Lee, and D.~Schritt,
  Phys.\ Lett.\ B {\bf 687} (2010) 248,
  arXiv:0804.1854 [hep-th].

\bibitem{Ahluwalia:2009rh}
  D.~V.~Ahluwalia, C.~Y.~Lee, and D.~Schritt,
  Phys.\ Rev.\ D {\bf 83} (2011) 065017,
  arXiv:0911.2947 [hep-ph].

\bibitem{Ahluwalia:2010zn}
  D.~V.~Ahluwalia and S.~P.~Horvath,
  JHEP {\bf 1011}, (2010) 078,
  arXiv:1008.0436 [hep-ph].

\bibitem{Dias:2010aa}
  M.~Dias, F.~de Campos, and J.~M.~Hoff da Silva,
  Phys.\ Lett.\ B {\bf 706} (2012) 352,
  arXiv:1012.4642 [hep-ph].

\bibitem{Lee:2012td}
  C.~Y.~Lee,
  Int.\ J.\ Mod.\ Phys.\ A {\bf 30} (2015) 1550048,
  arXiv:1210.7916 [hep-th].

\bibitem{Lee:2015jpa}
  C.~Y.~Lee,
  Int.\ J.\ Mod.\ Phys.\ A {\bf 31}, no. 35, 1650187 (2016)
  arXiv:1510.04983 [hep-th].

\bibitem{Fabbri:2010va}
  L.~Fabbri,
  Gen.\ Rel.\ Grav.\  {\bf 43} (2011) 1607,
  arXiv:1008.0334 [gr-qc].

\bibitem{Boehmer:2008rz}
  C.~G.~Boehmer,
  Phys.\ Rev.\ D {\bf 77} (2008) 123535,
  arXiv:0804.0616 [astro-ph].

\bibitem{Boehmer:2008ah}
  C.~G.~Boehmer and J.~Burnett,
  Phys.\ Rev.\ D {\bf 78} (2008) 104001,
  arXiv:0809.0469 [gr-qc].

\bibitem{Boehmer:2010ma}
  C.~G.~Boehmer, J.~Burnett, D.~F.~Mota, and D.~J.~Shaw,
  JHEP {\bf 1007} (2010) 053,
  arXiv:1003.3858 [hep-th].

\bibitem{Wei:2010ad}
  H.~Wei,
  Phys.\ Lett.\ B {\bf 695} (2011) 307,
  arXiv:1002.4230 [gr-qc].

\bibitem{Gredat:2008qf}
  D.~Gredat and S.~Shankaranarayanan,
  JCAP {\bf 1001} (2010) 008,
  arXiv:0807.3336 [astro-ph].

\bibitem{Basak:2014qea}
  A.~Basak and S.~Shankaranarayanan,
  JCAP {\bf 1505} (2015) no.05,  034,
  arXiv:1410.5768 [hep-ph].



\bibitem{Pereira:2014wta}
  S.~H.~Pereira, A.~P.~S.S., and J.~M.~Hoff da Silva,
  JCAP {\bf 1408} (2014) 020,
  arXiv:1402.6723 [gr-qc].


\bibitem{Pereira:2016emd}
  S.~H.~Pereira, A.~P.~S.S., J.~M.~Hoff da Silva, and J.~F.~Jesus,
  JCAP {\bf 1701} (2017) no.01,  055,
  arXiv:1608.02777 [gr-qc].

\bibitem{Pereira:2017efk}
  S.~H.~Pereira and T.~M.~$\text{Guimar}\tilde{\text{a}}\text{es}$,
  JCAP {\bf 09} (2017) 038,
  arXiv:1702.07385 [gr-qc].



\bibitem{Pereira:2017bvq}
  S.~H.~Pereira,
  \emph{Evolution of the universe driven by a mass dimension one fermion field},
  arXiv:1703.07636 [gr-qc].

\bibitem{HoffdaSilva:2009is}
  J.~M.~Hoff da Silva and R.~da Rocha,
  Int.\ J.\ Mod.\ Phys.\ A {\bf 24} (2009) 3227,
  arXiv:0903.2815 [math-ph].

\bibitem{HoffdaSilva:2016ffx}
  J.~M.~Hoff da Silva, C.~H.~Coronado Villalobos, R.~J.~Bueno Rogerio, and E.~Scatena,
  Eur.\ Phys.\ J.\ C {\bf 76} (2016) no.10,  563,
  arXiv:1608.05365 [hep-th].

\bibitem{daRocha:2011yr}
  R.~da Rocha, A.~E.~Bernardini, and J.~M.~Hoff da Silva,
  JHEP {\bf 1104} (2011) 110,
  arXiv:1103.4759 [hep-th].

\bibitem{Ablamowicz:2014rpa}
  R.~Ab{\l}amowicz, I.~Gon\c{c}alves, and R.~da Rocha,
  J.\ Math.\ Phys.\  {\bf 55} (2014) 103501,
  arXiv:1409.4550 [math-ph].

\bibitem{Rogerio:2016grn}
  R.~J.~Bueno Rogerio, J.~M.~Hoff da Silva, S.~H.~Pereira, and R.~da Rocha,
  Europhys.\ Lett.\  {\bf 113} (2016) no.6,  60001,
  arXiv:1603.09183 [hep-th].

\bibitem{Cavalcanti:2014wia}
  R.~T.~Cavalcanti,
  Int.\ J.\ Mod.\ Phys.\ D {\bf 23} (2014) no.14,  1444002,
  arXiv:1408.0720 [hep-th].

\bibitem{Cavalcanti:2014uta}
  R.~T.~Cavalcanti, J.~M.~Hoff da Silva, and R.~da Rocha,
  Eur.\ Phys.\ J.\ Plus {\bf 129} (2014) no.11,  246,
  arXiv:1401.7527 [hep-th].

\bibitem{Liu:2011nb}
  Y.~X.~Liu, X.~N.~Zhou, K.~Yang, and F.~W.~Chen,
  Phys.\ Rev.\ D {\bf 86} (2012) 064012,
  arXiv:1107.2506 [hep-th].

\bibitem{Jardim:2014cya}
  I.~C.~Jardim, G.~Alencar, R.~R.~Landim, and R.~N.~Costa Filho,
  Phys.\ Rev.\ D {\bf 91} (2015) no.4,  048501,
  arXiv:1411.5980 [hep-th].


\bibitem{Jardim:2014xla}
  I.~C.~Jardim, G.~Alencar, R.~R.~Landim, and R.~N.~Costa Filho,
  Phys.\ Rev.\ D {\bf 91} (2015) no.8,  085008,
  arXiv:1411.6962 [hep-th].

\bibitem{Dantas:2015mfi}
  D.~M.~Dantas, R.~da Rocha, and C.~A.~S.~Almeida,
  Europhys.\ Lett.\  {\bf 117} (2017) no.5,  51001,
  arXiv:1512.07888 [hep-th].


%
%
%
%
%
%
%
%
%
%
%
%
%
%
%
%
%
%
%
%
%
%
%
%
%
%
%
%
%
%
%
%
%
%
%
%
%
%
%
%
%
%
%
%
%
%
%
%
%
%
%
%
%
%
%
%
%
%
%
%
%
%
%
%
%
%
%
%
%
%
%
%
%
%
%
%
%
%
%
%
%
%
%
%
%
%
%
%
%
%
%
%
%
%
%
%
%
%
%
%
%
%
%
%
%
%
%
%
%
%
%
%
%
%
%
%
%
%
%
%

\end{thebibliography}

\end{document}